# Comments Regarding "On the Nature of Science"

Amy Courtney, PhD
Department of Physics, United States Military Academy, West Point, NY 10996
Amy_Courtney@post.harvard.edu

Michael Courtney, PhD
Ballistics Testing Group, P.O. Box 24, West Point, NY 10996
Michael_Courtney@alum.mit.edu

**Abstract:** An attempt to redefine science in the 21st century (BK Jennings, On the Nature of Science, Physics in Canada, 63(7) 2007) has abandoned traditional notions of natural law and objective reality, blurred the distinctions between natural science and natural history, elevated Occam's razor from an epistemological preference to a scientific principle, and elevated peer-review and experimental care as equals with repeatable experiment as arbiters of scientific validity. Our comments review the long-established axioms of the scientific method, remind readers of the distinctions between science and history, disprove the generality of Occam's razor by counter example, and highlight the risks of accepting additional scientific arbiters as equal to repeatable experiment.

"On the Nature of Science"[1] presents many valid points regarding the epistemology and practice of science in the 21st century, but there are several important incongruities.

Three axioms presupposed by the scientific method are realism (the existence of objective reality), the existence of observable natural laws, and the constancy of observable natural law.[2] Rather than depend on provability of these axioms, science depends on the fact that they have not been objectively falsified.

Occam's razor and related appeals to simplicity are epistemological preferences, not general principles of science. The general principle of science is that theories (or models) of natural law must be consistent with repeatable experimental observations. This principle rests upon the unproven axioms mentioned above. Occam's razor supports, but does not prove, these axioms.

There are many examples where Occam's razor would have picked the wrong theory given the available data. Simplicity principles are useful philosophical preferences for choosing a more likely theory from among several possibilities that are each consistent with available data. However, anyone invoking Occam's razor to support a scientific preference should be aware that future experiments may well falsify the model currently favored by Occam's razor. One accurate observation of a white crow falsifies the theory that "all crows are black." Likewise, a single instance of Occam's razor picking a wrong theory falsifies the razor as a general principle.

In addition, Occam's razor fails to acknowledge that if multiple models of natural law make exactly the same testable predictions, they are equivalent and there is no need for parsimony to choose one that is preferred. For example, Newtonian, Hamiltonian, and Lagrangian classical mechanics are equivalent. Which one of the three would be preferred by Occam's razor? Is this a justification for saying the other two are wrong? Likewise, how would advocates of simplicity principles arbitrate between wave and matrix formulations of quantum mechanics?

Defining science broadly as internally consistent and logical models supported by observation removes the requirement of repeatable experiment and broadens the realm of applicability from questions of natural law to questions of history and nearly every other question the human mind can devise. More importantly, it contains the implicit presupposition that observations that are limited in scope, audience, and duration are equally reliable as experimental observations that are repeatable a large number of times by any audience that bothers to exercise sufficient experimental care. One also wonders to what degree documented observations of other parties are admissible. Should less reliable observations (not repeatable but documented by others) have the same epistemological status as repeatable experiments?

The author describes the three-legged stool of scientific error correction as experimental care, reproducible experiment, and peer review. Like Occam's razor, peer review is a temporal expedient for judging scientific work at a given time. Peer review serves the temporal interests of publishers, funding agencies, and employers. The ultimate arbiter of scientific validity is repeatable experiment alone.

The three-legged stool of historical inquiry includes physical evidence (and scientific analysis thereof), documentary evidence, and eyewitness testimony.[3] When considering historical facts of legal significance, one is hard pressed to find many criminal convictions based only on scientific evidence without corroborating documentary or eyewitness testimony. The legal system recognizes that the availability and quality of evidence degrades with time. Increasing entropy guarantees that historical questions cannot be repeatedly tested as many times and by as many independent parties as questions of natural law.

Because models of natural law can be repeatedly tested by any audience that exercises sufficient experimental care, the reliability (or even the identity) of sources is less important in questions of natural law than in historical matters. Few scientists would object to acceptance of a new theory or even an anonymous experimental result if it was repeatable by independent sources. In contrast, the reliability of sources is critical in the believability of historical theories. For example, if those collecting evidence at a crime scene are found to be careless or dishonest in their handling of evidence, the prosecution stands very little chance of gaining a conviction based on the evidence collected, and the procedure simply cannot be repeated in a manner that counteracts the unreliability of the original sources.

The author shifts from a relatively narrow definition of science as reproducible experiment to "anything that can be observed" in order to make the case that supernatural claims can be investigated scientifically. However, this ignores the presupposition of constancy of natural law that underpins the demand of experimental repeatability as the ultimate arbiter in science. Since science presupposes the constancy of natural law in the demand of experimental repeatability, supernatural claims cannot be falsified scientifically without creating a circular argument. (In addition, most supernatural claims concern specific historical events rather than repeatable phenomena.) The objectivity of science rests in experimental repeatability.

Defining science as "observationally constrained model building" is barely more specific than defining science as "what scientists do." How far is this from defining sound science as "what scientists say" (with appropriate homage to peer review)? At this point, is science really a powerful, objective epistemology for exploring natural law, or have we merely replaced one set of authorities (the Catholic Church of the Middle Ages) with another (the scientists of the 21$^{st}$ century)?

We must not replace experimental repeatability with peer-reviewed observations as the ultimate arbiter of scientific validity. Only repeatable experimental results qualify as scientific observations. Observations of physical and documentary evidence of historical events do not warrant equal status with repeatable experiments.

Evolution as a model of natural history is widely accepted because many expectations have been fulfilled by later discoveries of pre-existing evidence. (Such later discoveries are better described as fulfilled expectations than specific predictions akin to Einstein's prediction of the bending of light by gravitation.) As a scientific theory of ongoing biological processes, evolution would be strengthened by more direct future observations of speciation in higher organisms via the expected mechanisms of natural selection.

"Predictive power" as an arbiter of science means the ability to predict the future course of systems under study, not merely the ability to predict future discoveries of pre-existing historical evidence. Valid models of natural law do not merely predict future discoveries of pre-existing evidence; they predict observations of future events.

Does failing to make an epistemological distinction between natural history and natural law really serve the interests of science? Does replacing the demand for predicting results of repeatable experiment with the much broader "observable support" better prepare today's student to become tomorrow's scientist?

Science needs objective criteria to rank the value of predictions and observations without the appeals to authority inherent in peer review or "scientific consensus." Observations that are experimentally repeatable should rank higher than historical observations whose repeatability is limited by increasing entropy. Specific predictions regarding future events should rank higher than expectations of future discoveries of pre-existing evidence. Thus, the science of natural law is inherently more objective than scientific descriptions of natural history.

What is the benefit of pretending that science provides the same high levels of certainty in historical theories of origins (species, universe, solar system) as the more objectively and repeatably testable quantum electrodynamics and classical mechanics (within their well-established areas of applicability)?

In conclusion, parsimony and failure to distinguish history from science can conspire to produce unwarranted levels of certainty. For example, consider an innocent man whose blood is found at a murder scene. In the absence of observable exculpatory evidence, the Jennings definition of science would seem to demand his conviction, regardless of plausible alternate explanations that might be offered, since alternate explanations are likely more complex than the theory that the defendant is guilty.

**Afterword:**
Jennings' reply fails to acknowledge objective reality regarding the laws of nature, but his scientific method implicitly demands the existence of objective reality regarding what scientific authors knew and when they knew it in order to make valid distinctions between what he refers to as "predictions" and "post-dictions."